\pdfoutput=1
\documentclass[11pt]{article}
\usepackage{amsmath}
\usepackage{graphicx,epstopdf}
\usepackage{cite}
\usepackage{array}
\usepackage{amsfonts}
\usepackage{graphicx}
\usepackage{float}
\usepackage[font={small}]{caption}
\usepackage{xcolor}
\usepackage[colorlinks=true,
            linkcolor=red,
            urlcolor=green,
            citecolor=blue]{hyperref}          

\catcode`@=12
\definecolor{RawSienna}{cmyk}{0,0.72,1,0.45}
\definecolor{dgreen}{rgb}{0.0,0.42,0.13}
\definecolor{darkblue}{rgb}{0.0, 0.0, 0.55}
\definecolor{cornellred}{rgb}{0.7, 0.11, 0.11}
\definecolor{calpolypomonagreen}{rgb}{0.08, 0.5, 0.5}

\newcommand{\B}{\color{darkblue}}

\def\beq{\begin{equation}}
\def\eeq{\end{equation}}
\def\bea{\begin{eqnarray}}
\def\eea{\end{eqnarray}}
\makeatletter
\@addtoreset{equation}{section}

\begin{document}
\title{\LARGE \bf Extended scaling and residual flavor symmetry in the neutrino Majorana mass matrix }
\author{{\bf Rome Samanta$^1$\footnote{ rome.samanta@saha.ac.in}, Probir Roy$^2$\footnote{probirrana@gmail.com}, Ambar Ghosal$^1$\footnote{ambar.ghosal@saha.ac.in},}\\
1. Astroparticle Physics and Cosmology Division\\Saha Institute of Nuclear Physics, HBNI,
  Kolkata 700064, India\\2. Center for Astroparticle Physics and Space Science \\ Bose Institute, Kolkata 700091, India} 
\maketitle
\begin{abstract}
The residual symmetry approach, along with a complex extension for some flavor invariance,  is a powerful tool to uncover the flavor structure of the $3\times3$ neutrino Majorana mass matrix $M_\nu$ towards gaining insights into neutrino mixing. We utilize this to propose a complex extension of the real scaling ansatz for $M_\nu$ which was introduced some years ago. Unlike the latter, our proposal allows a nonzero mass for each of the three light neutrinos as well as a nonvanishing $\theta_{13}$. The generation of light neutrino masses via type-I seesaw mechanism is also demonstrated. A major  result of this scheme is that leptonic Dirac CP-violation must be maximal while atmospheric neutrino mixing need not  be exactly maximal. Moreover, each of the two allowed Majorana phases, to be probed by the search for nuclear $0\nu \beta\beta$   decay, has to be at one of its two CP-conserving values. There are other interesting consequences such as the allowed occurrence of a normal mass ordering which is not favored by the real scaling ansatz. Our predictions will be tested in ongoing and future neutrino oscillation experiments at T2K, NO$\nu$A and DUNE.
\end{abstract}
\section{Introduction}\label{s1}
The masses and mixing properties of the three light neutrinos are beginning to get pinned down. Though the precise mass values are still unknown, upper limits on them have been pushed down to fractions of electron volts. Furthermore, it is already known that at least one of the neutrinos must be heavier than about 50 meV. Additionally, the three angles which describe their mixing have become reasonably well-known with $\theta_{12}\sim 34^o$, $\theta_{23}\sim 45^o$ and $\theta_{13}\sim 8^o$. Understanding this mixing phenomenon (with one small and two large angles) has emerged as a major challenge. As ongoing experiments feed in more and more information on neutrino masses and mixing, the flavor structure of the $3\times 3$ neutrino mass matrix $M_\nu$ is  being slowly uncovered. Many of its features still remain unknown nonetheless and continue to intrigue theoretical investigators. (Uptodate overviews of these issues and their investigations along with original references may be found in the two review articles quoted in Ref.\cite{King:2015aea}). Especially tantalizing is the predicted  phenomenon of leptonic CP-violation which likely to have implications for leptogenesis \cite{Hagedorn:2016lva}. As yet, there is no statistically reliable definitive experimental result on leptonic CP-violation. However, hints of a near-maximal CP-violation, with the phase $\delta$ being $\simeq$ $3\pi/2$, have emerged from results reported by the T2K \cite{Abe:2015awa}, NO$\nu$A \cite{Bian:2015opa} and Super-Kamiokande \cite{skk} experiments. Similarly, a recent global analysis \cite{Gonzalez-Garcia:2015qrr} of all neutrino data is hinting at a nonmaximal value of $\sin^2 2\theta_{23}$. Another yet unresolved question of great interest is that of  neutrino mass ordering : normal vs. inverted. In addition, one would like to know if the three neutrinos are Majorana or Dirac particles $-$ to be presumably determined by a future observation of nuclear $0 \nu \beta \beta$ decay \cite{Dell'Oro:2015oey}.

Let us start with the minimal supposition that there are only three light and flavored left-chiral neutrinos and that they are Majorana in character. The neutrino mass term in the Lagrangian density now reads
\bea
-\mathcal{L}_{mass}^\nu= \frac{1}{2}\bar{\nu_l^C} (M_\nu)_{lm}\nu_m + h.c. \label{lag}
\eea 
with $\nu_l^C=C\bar{\nu_l}^T$ and the subscripts $l,m$ spanning the lepton flavor indices $e$, $\mu$, $\tau$. $M_\nu$ is a complex symmetric matrix ($M_\nu^*\neq M_{\nu}=M_\nu^T$) which can be put into a diagonal form by a similarity transformation with a unitary matrix $U$:
\bea
U^T M_\nu U=M_\nu^d \equiv \rm diag\hspace{1mm}(m_1,m_2,m_3).\label{e0}
\eea  
Here ${\rm m}_i\hspace{1mm}(i=1,2,3)$ are real and positive masses. We choose to work in a Weak Basis \cite{branco} in which the charged lepton mass matrix is diagonal with real and positive elements, i.e. $M_l= \rm diag.\hspace{1mm}(m_e,m_\mu,m_\tau)$ and the unphysical phases of $U$ are absorbed into the neutrino fields. Now
\bea
U=U_{PMNS}\equiv 
\begin{pmatrix}
c_{1 2}c_{1 3} & e^{i\frac{\alpha}{2}} s_{1 2}c_{1 3} & s_{1 3}e^{-i(\delta - \frac{\beta}{2})}\\
-s_{1 2}c_{2 3}-c_{1 2}s_{2 3}s_{1 3} e^{i\delta }& e^{i\frac{\alpha}{2}} (c_{1 2}c_{2 3}-s_{1 2}s_{1 3} s_{2 3} e^{i\delta}) & c_{1 3}s_{2 3}e^{i\frac{\beta}{2}} \\
s_{1 2}s_{2 3}-c_{1 2}s_{1 3}c_{2 3}e^{i\delta} & e^{i\frac{\alpha}{2}} (-c_{1 2}s_{2 3}-s_{1 2}s_{1 3}c_{2 3}e^{i\delta}) & c_{1 3}c_{2 3}e^{i\frac{\beta}{2}} 
\end{pmatrix}\label{eu}
\eea
with $c_{ij}\equiv\cos\theta_{ij}$, $s_{ij}\equiv\sin\theta_{ij}$ and $\theta_{ij}=[0,\pi/2]$. CP-violation enters through nontrivial values of the Dirac phase $\delta$ and of the Majorana phases $\alpha,\beta$  with $\delta,\alpha,\beta=[0,2\pi]$. We follow the PDG convention \cite{PDG} on these angles and phases except that we denote the Majorana phases by $\alpha$ and $\beta$. In principle there could also be a phase matrix with $U_{PMNS}$ if we work in a Weak Basis where  $M_l$ is diagonal but where the unphysical phases are not absorbed in the neutrino fields.  It is  demonstrated later that even if we include the unphysical phase matrix, our result remains the same which is obvious, since physical results are basis independent.

Quite a few different hypotheses have been advanced over several decades on the flavor structure of $M_\nu$, as reviewed in the first article of Ref.\cite{King:2015aea}. We zero in on an ansatz made some years ago \cite{models} that we call Simple Real Scaling (SRS). This posits the relations
\bea
\frac{(M_{\nu}^{SRS})_{e\mu}}{(-M_{\nu}^{SRS})_{e\tau}}=\frac{(M_{\nu}^{SRS})_{\mu\mu}}{(-M_{\nu}^{SRS})_{\mu\tau}}=\frac{(M_{\nu}^{SRS})_{\tau\mu}}{(-M_{\nu}^{SRS})_{\tau\tau}}=k,\label{e1}
\eea
where $k$ is a real and positive dimensionless scaling factor. It is straightforward to induce from (\ref{e1}) the form of the neutrino Majorana mass matrix:
\bea
M_{\nu}^{SRS}=\begin{pmatrix}
X&-Yk& Y\\-Yk&Zk^2&-Zk\\Y&-Zk&Z
\end{pmatrix}.\label{e2}
\eea
Here $X$, $Y$, $Z$  are complex  mass dimensional quantities that are a priori unknown.  We  consistently denote complex (real) quantities by capital (small) letters throughout. We have chosen appropriate negative signs in (\ref{e1}) and (\ref{e2}) to be in conformity with the PDG convention \cite{PDG} on the form of $U_{PMNS}$ that emerges from (\ref{e2}). It was pointed out by Mohapatra and Rodejohann \cite{models} that - in the basis where the charged lepton mass matrix is diagonal - (\ref{e2}) can be realized from the larger symmetry group $D_4\times \mathbb{Z}_2$. This ansatz of Simple Real Scaling  led to a sizable body of research \cite{scaling}. But it predicts  a vanishing $s_{13}$ (and hence no measurable leptonic Dirac CP-violation) as well as an inverted neutrino mass hierarchy (i.e. $m_{2,1}>\hspace{1mm}m_3$) with $m_3=0$. While the latter result is still allowed within current experimental bounds, a null value of $s_{13}$ has been  ruled out at more than 10$\sigma$ \cite{th13}. Thus SRS, as it stands, has to be abandoned.

We want to consider an extended version of (\ref{e2}) which allows a nonvanishing $s_{13}$. To this end, we employ the method of complex  extension which in turn is based on the idea of the residual symmetry $\mathbb{Z}_2 \times \mathbb{Z}_2$\cite{Lam} of $M_\nu$. This is explained in Sec. \ref{s3} below. As detailed in the subsequent Sec. \ref{s4}, the complex  extension (CES for Complex Extended Scaling) leads to the neutrino mass matrix
\bea
 M_\nu^{CES} = \begin{pmatrix}
 x& -y_1 k +iy_2k^{-1}&y_1+iy_2\\
 -y_1 k +iy_2k^{-1}&z_1-wk^{-1}(k^2-1)-iz_2&w-iz_2(2k)^{-1}(k^2-1)\\
 y_1+iy_2 &w-iz_2(2k)^{-1}(k^2-1)&z_1+i z_2
\end{pmatrix}. \label{e3}
\eea 
Here $x$, $y_{1,2}$, $z_{1,2}$ and $ w$ are real mass dimensional quantities that are a priori unknown. It will be shown that $M_\nu^{CES}$ of (\ref{e3}) can accommodate a nonzero value for each  of $m_1$, $m_2$, $m_3$  and can fit the extant data on $\Delta m_{21}^2\equiv m_2^2-m_1^2$, $|\Delta m_{32}^2|\equiv|m_3^2-m_2^2|$ as well as on $\theta_{12}$ and $\theta_{13}$. The relation $\tan \theta_{23}=k^{-1}$ is a consequence so that the presently allowed range of $\tan\theta_{23}$ around unity would yield the permitted domain of the variation of the scaling parameter $k$ close to 1. Furthermore, (\ref{e3}) leads to the result that $\alpha,\beta$ = $0$ or $\pi$, i.e. there is no Majorana CP-violation, and the verifiable/falsifiable prediction that $\cos\delta=0$, i.e. leptonic Dirac CP-violation is maximal. We have no statement on the sign of $\sin \delta$ so that $\delta$ can be either $\pi/2$ or $3\pi/2$. Furthermore, we show that a normal mass ordering (with $m_{2,1}<m_3$) is allowed in addition to an inverted one ($m_{2,1}>m_3$) in the parameter space of the model.

The rest of the paper is organized as follows. In Sec.\ref{s2} we elucidate the meaning of the residual $ \mathbb{Z}_2\times \mathbb{Z}_2$ discrete symmetry of $M_\nu$ in terms of its invariance under two separate similarity transformations. Simple real scaling and its real generalization are discussed in Sec.\ref{s3}. Sec.\ref{s4} contains a presentation of the procedure of  complex  extension; this is first illustrated for $\mu\tau$ interchange symmetry and then applied to the scaling transformation to lead to the proposed $M_\nu^{CES}$ of (\ref{e3}) as well as its main consequences, namely $\tan\theta_{23}=k^{-1}$ and $\cos\delta=0$ plus the allowed occurrence of a normal mass ordering. The origin of the neutrino mass matrix $M_\nu^{CES}$ in our scheme from type-I seesaw mechanism is shown in Sec.\ref{seesaw}. Detailed phenomenological implications of $M_\nu^{CES}$ are worked out numerically in Sec.\ref{s5} and fitted with the current data yielding various $3\sigma$-allowed regions in the parameter space; the application of our results to forthcoming experiments on nuclear $0\nu\beta\beta$ decay and neutrino oscillations is also discussed in the same section. Sec.\ref{s6} summarizes our conclusions.
\section{Meaning of residual flavor symmetry of $M_\nu$}\label{s2}
It would be useful to focus on the feature \cite{Lam} of $M_\nu$ that it has a residual (sometimes called `remnant'\cite{ding}) $\mathbb{Z}_2\times \mathbb{Z}_2$ flavor symmetry and at the same time review the representation content of the latter. Such an exercise will enable us to set up the theoretical machinery needed to apply the idea to Simple Real Scaling. In addition, this will lead us to its real generalization as well as to its complex  extension.

Let $G$ be a generic $3\times3$ unitary matrix representation of some horizontal symmetry of $M_\nu$ effected through the similarity transformation
\bea
G^T M_\nu G=M_\nu \label{e4}. 
\eea 
Eqs. (\ref{e0}) and (\ref{e4}) lead to the conclusion that the unitary matrix $U^{\prime}\equiv GU$ also puts $M_\nu$ into a diagonal form by a similarity transformation, i.e. $U^{\prime T} M_\nu U^{\prime}=M_\nu^d$. It can then be shown \cite{Lam} that, if $m_1$, $m_2$ and $m_3$ are nondegenerate, $G$ has eigenvalues $\pm 1$ and is diagonalized by $U$. Thus 
\bea
GU=Ud,\label{e5}\\
d^2=I.\label{e6}
\eea
Here $d$ is a $3\times3$ diagonal matrix in flavor space with $d_{lm}= \pm\delta_{lm}$. There are eight possible distinct forms for $d$. Two of these are trivial $\textendash$ being the unit and the negative unit matrices. Of the remaining six, three are negatives of the other three. Finally, we have three $G_a$'s (a = 1, 2, 3) but it is sufficient to consider  any two of those as independent on account of the relation $G_a=\epsilon_{abc}G_bG_c$. The two independent $G_a$'s (chosen here as $G_{2,3}$) are representations of a residual $\mathbb{Z}_2 \times \mathbb{Z}_2$ symmetry in the Majorana mass term of the neutrino Lagrangian. It follows from  (\ref{e5}) and (\ref{e6}) that
\bea
G^2=I,\label{e7}\\
\rm det\hspace{1mm} G=\pm 1. \label{e8}
\eea
The eigenvalue equation (\ref{e5}) needs to be considered for the two independent $d$\hspace{.5mm}$^,$s, i.e. $d_2$ and $d_3$, corresponding respectively to $G_2$ and $G_3$. Suppose we choose
\bea
d_2=\rm diag\hspace{1mm} (-1,1,-1), \label{e9}\\
d_3=\rm diag\hspace{1mm} (-1,-1,1) \label{e10}
\eea
for $\rm det$ $G=$1. (The choice for the case $\rm det$ $G=-1$ is a trivial extension with $-d_2$ and $-d_3$.) Now
\bea 
G_{2,3}=Ud_{2,3}U^{\dagger}\label{ge}
\eea
 can be obtained by use of the explicit form of $U$ as given in (\ref{eu}). For instance, let us consider the situation for $\mu\tau$ interchange symmetry \cite{mutau} which implies $\theta_{23}=\pi/4$ and $\theta_{13}=0$. Now we obtain 
\bea
G_2=\begin{pmatrix}
-\cos 2\theta_{12}&2^{-\frac{1}{2}}\sin 2\theta_{12}&-2^{-\frac{1}{2}}\sin 2\theta_{12}\\ 2^{-\frac{1}{2}}\sin 2\theta_{12}&-\frac{1}{2}(1-\cos 2\theta_{12})&-\frac{1}{2}(1+\cos 2\theta_{12})\\ -2^{-\frac{1}{2}}\sin 2\theta_{12}&-\frac{1}{2}(1+\cos 2\theta_{12})&-\frac{1}{2}(1-\cos 2\theta_{12})
\end{pmatrix},
G_3^{\mu\tau}=
\begin{pmatrix}
-1&0&0\\0&0&1\\0&1&0
\end{pmatrix}.\label{e11}
\eea
The above $G_3$ explicitly implements $\mu\tau$ interchange in the neutrino flavor basis and hence has been labeled with the superscript $\mu\tau$. Thus one can now identify one of the two residual $\mathbb{Z}_2$\hspace{0.5mm}$^{,}$s as $\mathbb{Z}_2^{\mu \tau}$.  The full residual symmetry in this case is $\mathbb{Z}_2\times \mathbb{Z}_2^{\mu\tau}$. 
Our aim would be to undertake a similar task with scaling symmetry in obtaining a $\mathbb{Z}_2^{scaling}$. It may be mentioned that some authors \cite{Grimus:2004cc} have generalized $G_3^{\mu\tau}$ to
\bea
G_3^{G\mu\tau}=\begin{pmatrix}
-1&0&0\\0&\cos 2\theta_{23} & \sin 2\theta_{23}\\0&\sin 2\theta_{23}& \cos 2\theta_{23}
\end{pmatrix}
\eea 
which can accommodate an arbitrary $\theta_{23}$ but still has $\theta_{13}=0$. A somewhat different use of the residual symmetry approach with another pair of $\mathbb{Z}_2$\hspace{0.5mm}$^{,}$s was made in Ref.\cite{dicus}.

 A comment on the use of the residual $\mathbb{Z}_2 \times \mathbb{Z}_2$ symmetry would be in order. One could start from any arbitrary ansatz on $U_{PMNS}$, reconstruct the residual $\mathbb{Z}_2 \times \mathbb{Z}_2$ symmetry and work out the consequences. However, the $\mathbb{Z}_2 \times \mathbb{Z}_2$ symmetry emerging from an arbitrary ansatz may not follow from a larger symmetry group or have some deeper flavor meaning. The SRS ansatz has been shown to follow \cite{models} from a larger flavor symmetry group $D_4\times \mathbb{Z}_2$.
\section{Simple Real Scaling and its real generalization}\label{s3}
 Simple Real Scaling and the corresponding $M_\nu^{SRS}$, cf. (\ref{e2}), were already introduced in Sec.\ref{s1}. It is evident from (\ref{e2}) that the latter has a vanishing determinant, i.e. one null eigenvalue. The corresponding eigenvector, given that $\theta_{12}$ and $\theta_{23}$ are known to be hugely nonzero, can be identified only with the third column of $U^{SRS}$ and written as 
\bea
C_3^{SRS}=\begin{pmatrix}

0\\ (1+k^2)^{-\frac{1}{2}}e^{i\frac{\beta}{2}} \\ k(1+k^2)^{-\frac{1}{2}}e^{i\frac{\beta}{2}}
\end{pmatrix}.\label{e12}
\eea
Two immediate consequences are that $m_3=0$, i.e. the neutrino mass ordering is inverted ($m_{2,1}>m_3$), and $\theta_{13}=0$. The full $U^{SRS}$ can be written with an undetermined angle $\theta_{12}$ and the corresponding $c_{12}$, $s_{12}$ as
\bea
U^{SRS}=\begin{pmatrix}
c_{12}&s_{12}e^{i\frac{\alpha}{2}}&0\\
-k(1+k^2)^{-\frac{1}{2}}s_{12}&k(1+k^2)^{-\frac{1}{2}}c_{12}e^{i\frac{\alpha}{2}}&(1+k^2)^{-\frac{1}{2}}e^{i\frac{\beta}{2}}\\
(1+k^2)^{-\frac{1}{2}}s_{12}&-(1+k^2)^{-\frac{1}{2}}c_{12}e^{i\frac{\alpha}{2}}&k(1+k^2)^{-\frac{1}{2}}e^{i\frac{\beta}{2}}
\end{pmatrix}.\label{e13}
\eea
A comparison between (\ref{eu}) and (\ref{e13}) immediately yields
\bea
\tan \theta_{23}=k^{-1}.\label{e14}
\eea
As we shall see, (\ref{e14}) is going to survive both the real generalization and the complex CP-transformed extension of SRS. 

An expression for $G_3^{scaling}$ as a representation for $\mathbb{Z}_2^{scaling}$ can now be derived by use of (\ref{ge}) . On utilizing $U^{SRS}$ from (\ref{e13}) and $d_3$ from (\ref{e10}), we have 
\bea
G_3^{scaling}=
\begin{pmatrix}
-1&0&0\\0&(1-k^2)(1+k^2)^{-1}&2k(1+k^2)^{-1}\\
0&2k(1+k^2)^{-1}&-(1-k^2)(1+k^2)^{-1}
\end{pmatrix}=(G_3^{scaling})^T.\label{e15}
\eea
The $\mathbb{Z}_2^{scaling }$ symmetry of $M_\nu^{SRS}$ ensures that
\bea
(G_3^{scaling})^T M_\nu^{SRS} G_3^{scaling}=M_\nu^{SRS}.\label{e16}
\eea
It may be noted that (\ref{e16}) does not lead uniquely to the form (\ref{e2}). Further, while the form of $G_3^{scaling}$ follows uniquely from $U^{SRS}$ of (\ref{e13}) via the relation been $G_3$ and $d_3$, the reverse is not the case. Indeed, though the third column of $U$, reconstructed from $G_3^{scaling}$, must be $C_3^{SRS}$ of (\ref{e12}) since $(d_3)_{33}=1$, its first two columns could be an arbitrary orthogonal pair. That occurs because of the degeneracy of the (1,1) and (2,2) elements in $d_3$. The full residual symmetry of $M_\nu^{SRS}$ is $\mathbb{Z}_2^{k}\times \mathbb{Z}_2^{scaling}$, where a representation for $\mathbb{Z}_2^k$ is $G_2^k=U^{SRS}d_2 U^{SRS\dagger}$. Explicitly,
\bea
G_2^k=\begin{pmatrix}
-\cos 2\theta_{12} & k (1+k^2)^{-1}\sin 2\theta_{12}& -(1+k^2)^{-1}\sin 2\theta_{12}\\
 k(1+k^2)^{-1}\sin 2\theta_{12}&-k^2(1+k^2)^{-1}(1-\cos2\theta_{12})&-k(1+k^2)^{-1}(1+\cos2\theta_{12})\\
  -(1+k^2)^{-1}\sin 2\theta_{12}&-k(1+k^2)^{-1}(1+\cos2\theta_{12})&-(1+k^2)^{-1}(k^2-\cos 2\theta_{12})
\end{pmatrix}\label{e17}
\eea
which obeys 
\bea
(G_2^{k})^T M_\nu^{SRS} G_2^{k}=M_\nu^{SRS}.
\eea

A good check is that, for $k=1$, the scaling procedure just reduces to $\mu\tau$ interchange with the additional constraint $M^\nu_{\mu\mu}=M^\nu_{\mu\tau}$. But the point of real interest is that $M_\nu^{SRS}$ of (\ref{e2}) is not the most general form obeying (\ref{e16}). The latter may be worked out to be 
\bea
M_\nu^{GRS}=\begin{pmatrix}
x&-Yk&Y\\-Yk&Z-Wk^{-1}(k^2-1)&W\\Y&W&Z
\end{pmatrix},\label{e19}
\eea
where $W$ is another a priori unknown mass dimensional complex quantity. We call this form of $M_\nu$ the Generalized Real Scaling ansatz and denote it by the superscript $GRS$. Evidently, the specific choice $W=-Zk$ reduces $M_\nu^{GRS}$ to $M_\nu^{SRS}$. The neutrino mass matrix $M_\nu^{GRS}$ of (\ref{e19}) has interesting properties. For one thing, it has a determinant which does not appear to vanish. Therefore, we take all neutrino masses to be nonzero and can accommodate a nonzero $m_3$ and in principle a normal mass ordering with $m_{2,1}<m_3$. However, being invariant under a similarity transformation by $G_3^{scaling}$ of (\ref{e15}), the third column of the corresponding $U^{GRS}$ is constrained to be $C_3^{SRS}$ of (\ref{e12}). Consequently, one obtains a vanishing $\theta_{13}$ which is now experimentally known to be nonzero. Thus $M_\nu^{GRS}$ of (\ref{e19}) is unacceptable. A more extended version of scaling in the neutrino mass matrix is needed to describe nature. This is what will be provided in the next section.
\section{Complex  extension of scaling ansatz}\label{s4}
It would be useful to first recall how the complex  extension of $\mu\tau$ interchange symmetry was originally made \cite{mutau}. The $\mu\tau$ interchange invariant $M_\nu^{\mu\tau}$ obeys the condition
\bea
(G_3^{\mu\tau})^T M_\nu^{\mu\tau}G_3^{\mu\tau}=M_\nu^{\mu\tau}\label{e20}
\eea
with $G_3^{\mu\tau}$ given by (\ref{e11}). Eq. (\ref{e20}) forces $M_{\nu}^{\mu\tau}$ to have the form
\bea
M_\nu^{\mu\tau}=\begin{pmatrix}
A&B&B\\B&C&D\\B&D&C
\end{pmatrix},\label{e21}
\eea
with $A$, $B$, $C$, $D$ as mass dimensional complex quantities. It is well-known that (\ref{e21}) leads to $\theta_{13}=0$ and cannot be accepted as it stands. 

Grimus and Lavoura made an alternative proposal, namely the complex-extended invariance relation
\bea
(G_3^{\mu\tau})^T M_\nu G_3^{\mu\tau}=M_\nu^{*}.\label{e22}
\eea
This was justified \cite{mutau} by means of a non-standard CP-transformation \cite{gencp} on the $\nu_e$ field  which is generally represented as\footnote{It is a theoretically interesting question whether such an extended CP-invariance can arise from an automorphism of a larger flavor symmetry like in the top-down approach of Ref.\cite{Feruglio:2012cw}. But we do not explore this possibility here.}
\bea
\nu_{L\alpha }\rightarrow i G_{\alpha \beta}\gamma^0 \nu_{L\beta}^C \label{CPtr}
\eea 
with $G_{\alpha \beta}$ as the matrix element of the flavor symmetry. Eq. (\ref{CPtr}) along with (\ref{lag}) leads to (\ref{e22}) if $G_{\alpha \beta}$ is considered as $G_3^{\mu \tau}$. Suffice it to say that (\ref{e22}) leads to a complex-extended $\mu\tau$ ($CE\mu\tau$) symmetric form of $M_\nu$:
\bea
M_\nu^{CE\mu\tau}=\begin{pmatrix}
a&B&B^*\\
B&C&d\\
B^*&d&C^*
\end{pmatrix},\label{e23}
\eea
where $a$, $d$ are real and $B$, $C$ are complex mass dimensional quantities in general. Once again, since the determinant does not vanish, we take all neutrino masses to be nonzero. The observable consequences of (\ref{e23}) are: $\theta_{23}=\pi/4$, $\cos\delta=0$, $\alpha$, $\beta$ = 0 or $\pi$ while $\theta_{13}$ is in general nonzero. A further extension of this approach has recently been made \cite{ding,chen} allowing  nonmaximal values for $\theta_{23}$ and  Dirac CP-violation. 

We have derived (\ref{e14}), i.e. $\tan\theta_{23}=k^{-1}$, so that atmospheric neutrino mixing is not forced to be strictly maximal.  On the other hand, the observed fact that $\tan \theta_{23}$ is not far from unity implies that so is $k$. Our proposed relation, in place  of (\ref{e22}), is
\bea
(G_3^{scaling})^T M_\nu G_3^{scaling}=M_\nu^*, \label{e24}
\eea
with $G_3^{scaling}$ as given in (\ref{e15}) and, as stated earlier, in the basis where the charged lepton mass matrix is diagonal and positive. The general form of $M_\nu^{CES}$, as given in (\ref{e3}), follows in consequence. It is important to note that $M_\nu^{CES}$ of (\ref{e3}) has a structure that is quite different from that of either $M_\nu^{SRS}$ of (\ref{e2}) or $M_{\nu}^{GRS}$ of (\ref{e19}). If all imaginary parts in $M_{\nu}^{CES}$ are set equal to zero, a form similar to that of $M_\nu^{GRS}$ is recovered but with all real entries while those in $M_{\nu}^{GRS}$ of (\ref{e19}) are in general complex. Therefore, no special choice in $M_\nu^{CES}$ can yield $M_\nu^{SRS}$ or $M_\nu^{GRS}$ in their respective generalities.

Grimus and Lavoura \cite{mutau} had proved a corollary of complex-extended invariance. This can be stated with respect to a relation such as (\ref{e22}) or (\ref{e24}) as 
\bea
G_3^{scaling}U^*=U\tilde{d} \label{e25}
\eea 
with $\tilde{d}$ as a diagonal matrix. Once again, $\tilde{d}_{lm}=\pm \hspace{1mm}\delta_{lm}$ if the neutrino masses $m_1$, $m_2$, $m_3$ are all nonzero and nondegenerate. The key difference between (\ref{e25}) and (\ref{e5}) is the complex conjugation of $U$ in the LHS. Let us take 
\bea
\tilde{d}= \rm diag\hspace{1mm} (\tilde{d_1},\tilde{d_2},\tilde{d_3})\label{26}
,\eea
where each $\tilde{d}_i$ $(i=1,2,3)$ can be $+1$ or $-1$. With $G_3=G_3^{scaling}$, (\ref{e25}) can be written explicitly :
\bea
\begin{pmatrix}
-(U_{e1}^{CES})^*&-(U_{e2}^{CES})^*&-(U_{e3}^{CES})^*\\
\frac{1-k^2}{1+k^2}(U_{\mu 1}^{CES})^*+\frac{2k}{1+k^2}(U_{\tau 1}^{CES})^*&\frac{1-k^2}{1+k^2}(U_{\mu 2}^{CES})^*+\frac{2k}{1+k^2}(U_{\tau 2}^{CES})^*&\frac{1-k^2}{1+k^2}(U_{\mu 3}^{CES})^*+\frac{2k}{1+k^2}(U_{\tau 3}^{CES})^*\\
\frac{2k}{1+k^2}(U_{\mu 1}^{CES})^*-\frac{1-k^2}{1+k^2}(U_{\tau 1}^{CES})^*&\frac{2k}{1+k^2}(U_{\mu 2}^{CES})^*-\frac{1-k^2}{1+k^2}(U_{\tau 2}^{CES})^*&\frac{2k}{1+k^2}(U_{\mu 3}^{CES})^*-\frac{1-k^2}{1+k^2}(U_{\tau 3}^{CES})^*
\end{pmatrix}\nonumber
\eea 

\bea
=\begin{pmatrix}
\tilde{d_1} U_{e 1}^{CES}&\tilde{d_2}U_{e 2}^{CES}&\tilde{d_3}U_{e 3}^{CES}\\
\tilde{d_1}U_{\mu 1}^{CES}&\tilde{d_2}U_{\mu 2}^{CES}&\tilde{d_3}U_{\mu 3}^{CES}\\
\tilde{d_1}U_{\tau 1}^{CES}&\tilde{d_2}U_{\tau 2}^{CES}&\tilde{d_3}U_{\tau 3}^{CES}
\end{pmatrix}. \label{e27}
\eea

It is evident from (\ref{e27}) that the choice $\tilde{d}_1=1$ leads to an imaginary $U_{e1}$ in contradiction with  the real (1,1) element of (\ref{eu}); this choice is hence excluded. Note that the choice of $U_{PMNS}$ in (\ref{eu}) is simply due to the choice of the Weak Basis where the neutrino fields are phase rotated. However, in Appendix \ref{A1}, we demonstrate that the physical results derived here are basis independent, i.e., the inclusion of an unphysical phase matrix does not impair our predictions.  There are now four permitted  cases $a$, $b$, $c$, $d$ with the following four combinations  allowed for $\tilde{d}$:
\bea
\tilde{d}_a&\equiv &\rm diag\hspace{1mm} (-1,1,1),\label{28a}\\
\tilde{d}_b&\equiv & \rm diag\hspace{1mm} (-1,1,-1),\label{28b}\\
\tilde{d}_c&\equiv & \rm diag\hspace{1mm} (-1,-1,1),\label{28c}\\
\tilde{d}_d& \equiv & \rm diag\hspace{1mm} (-1,-1,-1).\label{28d}
\eea
The above can be written compactly as
\bea
\tilde{d}_{a,b,c,d}= \rm diag\hspace{1mm}(-1,\eta,\xi)\\\label{29a}
\eta_{a,b}=1,\eta_{c,d}=-1,\\\label{29b}
\xi_{a,c}=1,\xi_{b,d}=-1.\label{29c}
\eea 
Comparing with (\ref{eu}), we obtain
\bea
e^{-i\alpha}=-\eta \label{30k}
\eea
\bea
e^{i(2\delta-\beta)}=-\xi. \label{30b}
\eea
Thus we are led to the result that $\alpha=\pi,0$ for $\eta=+1,-1$ respectively; in a similar manner $2\delta-\beta =\pi,0$ for $\xi=+1,-1$ respectively. We can derive from (\ref{e27}) altogether six  independent constraint conditions as linear relations among various elements of $U^{CES}$ and $(U^{CES})^*$. These are listed in Table \ref{t1}. 

More information is obtained by use of the explicit expressions of $U^{CES}_{l\alpha}$ from (\ref{eu}).
\begin{table}[!h]
\begin{center}
\caption{Constraint equations on elements of the mixing matrix}\label{t1}
 \begin{tabular}{|c|c|}
\hline
Element of $U^{CES}$ & Constraint condition\\
\hline
$\mu1$ & $2kU_{\mu 1}^{CES}=(1-k^2)U_{\tau 1}^{CES}-(1+k^2)(U_{\tau 1}^{CES})^*$\\
$\tau1$ & $2k U_{\tau 1}^{CES}=-(1-k^2)U_{\mu 1}^{CES}-(1+k^2)(U_{\mu 1}^{CES})^*$\\
$\mu 2$ & $2kU_{\mu 2}^{CES}=(1-k^2)U_{\tau 2}^{CES}+\eta(1+k^2)(U_{\tau 2}^{CES})^*$\\
$\tau 2$ & $2k U_{\tau 2}^{CES}=-(1-k^2)U_{\mu 2}^{CES}+\eta(1+k^2)(U_{\mu 2}^{CES})^*$\\
$\mu 3$ & $2kU_{\mu 3}^{CES}=(1-k^2)U_{\tau 3}^{CES}+\xi(1+k^2)(U_{\tau 3}^{CES})^*$\\
$\tau 3$ & $2k U_{\tau 3}^{CES}=-(1-k^2)U_{\mu 3}^{CES}+\eta(1+k^2)(U_{\mu 3}^{CES})^*$\\
\hline 
\end{tabular} 
\end{center} 
\end{table}
Consider the real and imaginary parts of the constraint condition on $U_{\tau 3}^{CES}$ given in the bottom line in Table \ref{t1}. Since $c_{13}$ is known to be nonzero, it can be canceled from both sides. Now, from the respective real and imaginary parts, we have the relations
\bea
2kc_{23}\cos\frac{\beta}{2}=[k^2(1+\xi)-1+\xi]s_{23}cos\frac{\beta}{2},\\\label{31a}
2kc_{23}\sin\frac{\beta}{2}=[k^2(1-\xi)-1-\xi]s_{23}\sin\frac{\beta}{2}.\label{31b}
\eea
Since $\xi^2=1$, the product of the above two equations leads to the result
\bea
\sin\beta = 0,
\eea 
or $\beta=0$ or $\pi$. There are now four options:

\bea
\beta &=&0, \xi=1\Rightarrow \tan\theta_{23}=k^{-1},\\
\beta &=&0, \xi=-1\Rightarrow \tan\theta_{23}=-k,\\
\beta &=&\pi, \xi=1\Rightarrow \tan\theta_{23}=-k,\\
\beta &=&\pi, \xi=-1\Rightarrow \tan\theta_{23}=k^{-1}.
\eea
 The option $\beta=0$, $\xi=1$ for cases $a$ and $c$, cf. (\ref{28a}) and (\ref{28c}), as well as $\beta=\pi$, $\xi=1$ for  cases $b$ and $d$, cf. (\ref{28b}) and (\ref{28d}), yield the scaling relation (\ref{e14}) while the other two options require $\tan\theta_{23}$ to equal $-k$. As will be shown below, the latter possibility is inconsistent with other constraint conditions. Our final result on the Majorana phases is that both $\alpha$ and $\beta$ are restricted to be $0$ or $\pi$. A combination of information from $0\nu\beta\beta$ decay, the cosmological upper bound on $\Sigma_im_i$ and the effective mass $\Sigma_i|U_{ei}|^2m_i$ measured in single $\beta$-decay is expected to experimentally constrain \cite{minakata} these phases.

To proceed further, consider the constraint condition on $U_{\tau 2}^{CES}$ given in the 4th line from the top of  Table \ref{t1}. The corresponding real and imaginary parts respectively yield
\bea
  &&2k[c_{12}s_{23}\cos\frac{\alpha}{2}+s_{12}s_{13}c_{23}\cos(\delta+\frac{\alpha}{2})]\nonumber\\
&&=[1-k^2-\eta(1+k^2)][c_{12}c_{23}\cos\frac{\alpha}{2}-s_{12}s_{13}s_{23}\cos(\delta+\frac{\alpha}{2})], \label{34a}\\
&&2k[c_{12}s_{23}\sin\frac{\alpha}{2}+s_{12}s_{13}c_{23}\sin(\delta+\frac{\alpha}{2})]\nonumber\\
&&=[1-k^2-\eta(1+k^2)][c_{12}c_{23}\sin\frac{\alpha}{2}-s_{12}s_{13}s_{23}\sin(\delta+\frac{\alpha}{2})]. \label{34b}
\eea
Let us now take the two cases at hand.\\
\textbf{Case 1:} $\eta=1$, $\alpha=\pi$\\
On utilizing that each of $s_{12}$, $s_{13}$ and $c_{23}$ is nonzero, one obtains from (\ref{34a}) and (\ref{34b}) the respective relations
\bea
&&(\tan\theta_{23}-k^{-1})\sin\delta=0, \label{35a}\\
&&c_{12}(\tan\theta_{23}-k^{-1})+s_{12}s_{13}(1+k^{-1}\tan\theta_{23})\cos\delta=0. \label{35b}
\eea
\textbf{Case 2:} $\eta=-1$, $\alpha=0$\\ 
It is easy to see that here one obtains the same pair of equations, namely (\ref{35a}) and (\ref{35b}), but in a reverse sequence. 

Eq. (\ref{35b}) has important implications. If $\tan\theta_{23}$ is put equal to $-k$, instead of $k^{-1}$, one is led to $c_{12}=0$ in contradiction with experiment \cite{Gonzalez-Garcia:2015qrr}. Therefore, the two options $\beta=0$, $\xi=1$ and $\beta=\pi$, $\xi=-1$ need to be retained with the other two options $\beta=0$, $\xi=-1$ and $\beta=\pi$, $\xi=1$  discarded. Now that $\tan\theta_{23}$ does equal $k^{-1}$, i.e. (\ref{e14}) holds, from (\ref{35b}) we have
\bea
\cos\delta=0, \label{36}
\eea
i.e. leptonic Dirac CP-violation is maximal with $\delta$ being either $\pi/2$ or $3\pi/2$. However, we are unable to distinguish between these two options since we have no statement on the sign of $\sin\delta$.
\begin{table}[!h]
\begin{center}
\caption{Predictions of the CP phases} \label{t2}
 \begin{tabular}{|c|c|c|c|} 
\hline 
$\tilde{d}$&$\alpha$ &$\beta$ &$ \cos \delta$ \\
\hline
$\tilde{d}_a=\rm diag \hspace{1mm}(-1,+1,+1)$&$\pi$&$0$&$0$\\
$\tilde{d}_b=\rm diag \hspace{1mm}(-1,+1,-1)$&$\pi$&$\pi$&$0$\\
$\tilde{d}_c=\rm diag \hspace{1mm}(-1,-1,+1)$&$0$&$0$&$0$\\
$\tilde{d}_d=\rm diag \hspace{1mm}(-1,-1,-1)$&$0$&$\pi$&$0$\\
\hline 
\end{tabular} 
\end{center} 
\end{table}
We have checked that (\ref{36}) consistently follows from the remaining four constraint equations of Table \ref{t1} and that no new condition emerges. Finally, we are left with four options, as shown in Table \ref{t2}. Each of these implies (\ref{36}), i.e. the maximality of leptonic Dirac CP violation which enters via $U_{PMNS}$.\\
\section{Origin of neutrino masses from type-I seesaw}\label{seesaw}
 In this section we discuss the realization of the complex extended scaling neutrino mass matrix $M_\nu^{CES}$ through the type-I seesaw mechanism via three heavy right-handed neutrino fields $N_{lR}\hspace{.5mm}(l=1,2,3)$ with a Majorana mass matrix $M_R$. We choose a  Weak Basis in which the charged lepton and the right-handed neutrino mass matrices are diagonal and nondegenarate. With $m_D$ as the Dirac mass matrix and $M_R=\rm diag\hspace{.5mm} (M_1,M_2,M_3)$, the neutrino mass terms read 
\bea
-\mathcal{L}_{mass}^{\nu,N}= \bar{N}_{lR} (m_D)_{l\alpha}L_{\alpha}+\frac{1}{2}\bar{N}_{lR}(M_R)_l \delta _{lm}N_{mR}^C + {\rm h.c}. \label{selag}
\eea
The effective light neutrino mass matrix is given by the standard seesaw relation
\bea
M_\nu = -m_D^TM_R^{-1}m_D. \label{swmnu}
\eea
We represent the G's, introduced earlier for left handed fields, generically by $G_L$ and define a corresponding $G_R$ for $N_R$. The residual CP transformations on the neutrino fields are defined as\cite{chen}
\bea
\nu_{L\alpha }\rightarrow i (G_L)_{\alpha \beta}\gamma^0 \nu_{L\beta}^C, \hspace{.5cm}
N_{R\alpha }\rightarrow i (G_R)_{\alpha \beta}\gamma^0 N_{R\beta}^C. \label{seCP}
\eea
The invariance of the mass terms of (\ref{selag}) under the CP transformations defined in (\ref{seCP}) leads to the relations
\bea
G_R^{\dagger}m_D G_L=m_D^*, \hspace{.3cm} G_R^{\dagger}M_R G_R^*=M_R^*. \label{trmdmr}
\eea
Eqs. (\ref{swmnu}) and (\ref{trmdmr}) together imply $G_L^TM_\nu G_L=M_\nu^*$. Now, specifying $G_L$ by $G_3^{scaling}$, we  obtain the key equation 
\bea
(G_3^{scaling})^T M_\nu G_3^{scaling}=M_\nu^*. 
\eea 
Since we choose the right handed neutrino mass matrix $M_R$ to be diagonal, the symmetry matrix $G_R$ is diagonal with entries $\pm1$, i.e.
\bea
G_R=\rm diag \hspace{1mm}(\pm1,\pm1,\pm1).
\eea
Hence there are eight different structures of $G_R$. Correspondingly, from the first relation of (\ref{trmdmr}), there are eight different structures of $m_D$. Unlike the complex transformations of $m_D$ and $M_R$ in (\ref{trmdmr}), we now have  real symmetry transformations  $G_R^{\dagger}m_D G_L=m_D$ and $ G_R^{\dagger}M_R G_R^*=M_R$. It can be shown by tedious algebra that, except for $G_R=\rm diag (-1,-1,-1)$, all other structures of $G_R$ are incompatible with  scaling symmetry, i.e. cannot generate $M_\nu^{GRS}$. Thus we take  $G_R=\rm diag (-1,-1,-1)$ as the only viable residual symmetry on the right-handed neutrino field. Now, $G_R^{\dagger}m_D G_L=m_D^*$ can be written as 
\bea
m_DG_L=-m_D^* \label{5p7}
\eea
which is basically a complex extension of the Joshipura-Rodejohann result $m_DG_L=-m_D$ \cite{scaling}.
In our context, (\ref{5p7}) can be rewritten as
\bea
 m_DG_3^{scaling}=-m_D^*. \label{mdeq}\
 \eea
 The most general $m_D$ that satisfies (\ref{mdeq}) is 
 \bea
 m_D^{CES}= \begin{pmatrix}
 a & b_1+ib_2 &-b_1/k+ib_2k\\
 e & c_1+ic_2 & -c_1/k+ic_2k\\
 f & d_1+id_2 & -d_1/k+id_2k\\
 \end{pmatrix},
 \eea
 where $a$, $b_{1,2}$, $c_{1,2}$, $d_{1,2}$, $e$ and $f$ are arbitrary real mass dimensional quantities. Using (\ref{swmnu}), $M_\nu^{CES}$ of (\ref{e3}) is obtained with the parameters as given in Table \ref{t}. Some detailed interesting consequences of $m_D^{CES}$, specifically with respect to leptogenesis, will be studied elsewhere.
\begin{table}[H]
\begin{center}
\caption{Parameters of $M_\nu^{CES}$ in terms of the parameters of $m_D$ and $M_R$} \label{t}
 \begin{tabular}{|c|} 
\hline 
$x=-(\frac{a^2}{M_1}+\frac{e^2}{M_2}+\frac{f^2}{M_3})$\\
$y_1=\frac{1}{k}(\frac{ab_1}{M_1}+\frac{ec_1}{M_2}+\frac{fd_1}{M_3})$\\
$y_2=k(\frac{ab_2}{M_1}+\frac{ec_2}{M_2}+\frac{fd_2}{M_3})$\\
$z_1=-\frac{1}{k^2}(\frac{b_1^2}{M_1}+\frac{c_1^2}{M_2}+\frac{d_1^2}{M_3})+k^2(\frac{b_2^2}{M_1}+\frac{c_2^2}{M_2}+\frac{d_2^2}{M_3})$\\
$z_2=\frac{2b_1b_2}{M_1}+\frac{2c_1c_2}{M_2}+\frac{2d_1d_2}{M_3}$\\
$w=\frac{1}{k}(\frac{b_1^2}{M_1}+\frac{c_1^2}{M_2}+\frac{d_1^2}{M_3})+k(\frac{b_2^2}{M_1}+\frac{c_2^2}{M_2}+\frac{d_2^2}{M_3})$\\
\hline
\end{tabular} 
\end{center} 
\end{table}

\section{Phenomenological constraints and consequences} \label{s5}
\begin{figure}[h!]
\hspace{.6cm}\includegraphics[scale=.4]{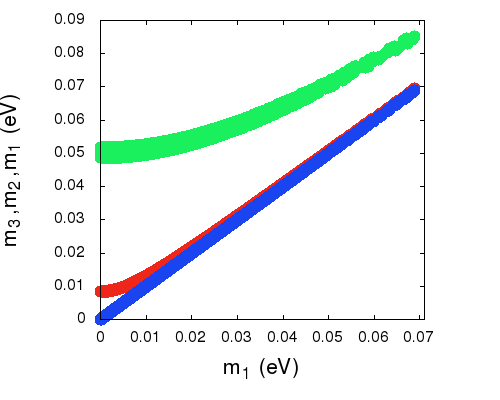}\hspace{1cm}\includegraphics[scale=.4]{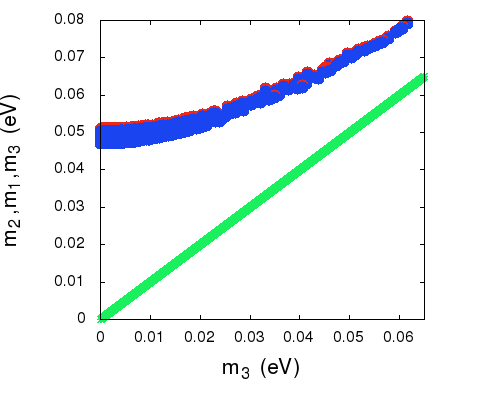}
\caption{Plots of the mass band for normal (left) and inverted (right) mass ordering. We have choosen to plot the lightest eigenvalue also in the ordinate to bring three mass bands together. Color code: green ($m_3$), red ($m_2$) and  blue ($m_1$). }\label{f1}
\end{figure}

We need to  numerically pin down the mass dimensional six real parameters $x$, $y_1$, $y_2$, $z_1$, $z_2$ and $w$ of $M_\nu^{CES}$ by inputting the 3$\sigma$ ranges of quantities measured in neutrino oscillation experiments. To that end, we use the values from a recent global analysis \cite{Gonzalez-Garcia:2015qrr}. In addition, we use the cosmological upper limit \cite{planck} of $0.23$ eV on the sum $m_1+m_2+m_3$ of the masses of the neutrinos. 
\begin{table}[h!]
\begin{center}
\caption{Input values used} \label{t3}
 \begin{tabular}{|c|c|c|c|c|c|} 
\hline 
$\theta_{12}$&$\theta_{23}$ &$\theta_{13}$ &$ \Delta m_{21}^2$&$|\Delta m_{31}^2|$&$\Sigma_i m_i$ \\
$\rm degrees$&$\rm degrees$ &$\rm degrees$ &$ 10^{-5}\rm eV^2$&$10^{-3} \rm (eV^2)$&$ \rm (eV) $ \\
\hline
$31.29-35.91$&$38.3-53.3$&$7.87-9.11$&$7.02-8.09$&$2.32-2.59$&$<0.23$\\
\hline
\end{tabular} 
\end{center} 
\end{table}
These input numbers are shown in Table \ref{t3}. In terms of output, we obtain the $3\sigma$ allowed intervals of the above mentioned six real parameters and from those the $3\sigma$ allowed ranges of the individual neutrino masses $m_1$, $m_2$, $m_3$. Both types of neutrino mass ordering, normal as well as inverted, are found to be allowed. All these values are listed in Tables \ref{t4} and \ref{t5} respectively  for the two separate categories of mass ordering.
\begin{table}[!h]
\begin{center}
\caption{Output values obtained for normal mass ordering with the best fit $m$'s given within brackets} \label{t4}
 \begin{tabular}{|c|c|c|c|c|c|} 
\hline 
$x$&$y_1$ &$y_2$ &$ z_1$&$z_2$&$w$ \\
$\rm (eV)$&$ \rm (eV)$ &$\rm (eV)$ &$\rm (eV)$&$ \rm (eV)$&$\rm (eV)$\\
\hline
$-0.20-+0.21$&$-0.12-+0.11$&$-0.05-+0.05$&$-0.17-+0.17$&$-0.18-+0.17$&$-0.16-+0.15$\\
\hline
 \multicolumn{2}{|c|}{\textbf{$m_1$}} & \multicolumn{2}{c|}{\textbf{$m_2$}} & \multicolumn{2}{c|}{\textbf{$m_3$}}\\
 \multicolumn{2}{|c|}{(eV)} & \multicolumn{2}{c|}{(eV)} & \multicolumn{2}{c|}{(eV)}\\
 \hline
 \multicolumn{2}{|c|}{$9.2\times10^{-5}-0.071$ (0.052) } & \multicolumn{2}{c|}{$0.01-0.077$ (0.054)} & \multicolumn{2}{c|}{$0.051-0.082$ (0.072)}\\
 \hline 
\end{tabular}  
\end{center} 
\end{table}
\begin{table}[!h]
\begin{center}
\caption{Output values obtained for inverted mass ordering with the best fit $m$'s given within brackets} \label{t5}
 \begin{tabular}{|c|c|c|c|c|c|} 
\hline 
$x$&$y_1$ &$y_2$ &$ z_1$&$z_2$&$w$ \\
$\rm (eV)$&$ \rm (eV)$ &$\rm (eV)$ &$\rm (eV)$&$ \rm (eV)$&$\rm (eV)$\\
\hline
$-0.44-+0.46$&$-0.16-+0.16$&$-0.14-+0.14$&$-0.01-+0.01$&$-0.01-+0.01$&$-0.05-+0.06$\\
\hline
\multicolumn{2}{|c|}{\textbf{$m_1$}} & \multicolumn{2}{c|}{\textbf{$m_2$}} & \multicolumn{2}{c|}{\textbf{$m_3$}}\\
 \multicolumn{2}{|c|}{(eV)} & \multicolumn{2}{c|}{(eV)} & \multicolumn{2}{c|}{(eV)}\\
 \hline
 \multicolumn{2}{|c|}{$0.049-0.079$ (0.068)} & \multicolumn{2}{c|}{$0.051-0.085$ (0.069)} & \multicolumn{2}{c|}{$8.2\times10^{-5}-0.068$ (0.048)}\\
 \hline 
\end{tabular}  
\end{center} 
\end{table}
We notice that, for both types of ordering, the neutrino masses become hierarchical, i.e. $m_{2,1}<<m_3$ for normal ordering and $m_{2,1}>>m_3$ for inverted ordering, for low values of the lightest neutrino mass. However they tend towards  quasi-degeneracy $m_1\sim m_2 \sim m_3$ as the latter increases to its permitted maximum value $\sim$ 0.07 eV. This is clear from the mass bands shown in Fig. \ref{f1}.\\

\noindent
{\it Neutrinoless double beta decay $0\nu\beta \beta$}\\

This is the lepton number violating process
\bea
(A,Z)\longrightarrow (A, Z+2)+2e^-
\eea
with no final state neutrinos. An  observation of the decay will confirm the Majorana nature of neutrinos which is yet to be established. The corresponding the half-life\cite{beta} is  given by
\bea
\frac{1}{T^{0\nu}_1/2}=G_{0\nu}|M_{0\nu}|^2 |M^\nu_{ee}|^2m_e^{-2}. 
\eea
where $G_{0\nu}$ is a phase space factor, $M_{0\nu}$ is the nuclear matrix element (NME), $m_e$ is the electron mass and finally $|M^\nu_{ee}|$ is the (1,1) element of $M^\nu$ which can also be written as $\Sigma_i m_iU_{ei}^2$. Following the PDG parametrization of the mixing matrix $U_{PMNS}$, one can write $M^\nu_{ee}$ as
\bea
M^\nu_{ee}=c_{12}^2c_{13}^2m_1+s_{12}^2c_{13}^2m_2e^{i\alpha}+s_{13}^2m_3e^{i(\beta-2\delta)}.
\eea
There are several ongoing experiments which have put significant upper limits on $|M^\nu_{ee}|$. Some recent experiments like KamLAND-Zen \cite{kam} and EXO \cite{exo} have improved this upper bound to $0.35$ eV. However, the most significant upper bound on $|M^\nu_{ee}|$ to date  is put by GERDA phase-I data \cite{gerda1} to be 0.22 eV; this is likely to be lowered by GERDA phase -II data\cite{gerda2} to around 0.098 eV. 

\begin{figure}[h!]
\includegraphics[scale=.4]{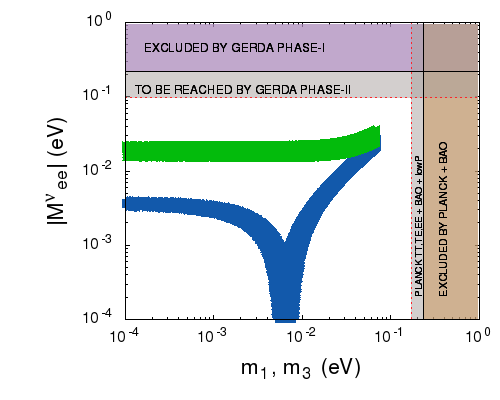} \includegraphics[scale=.4]{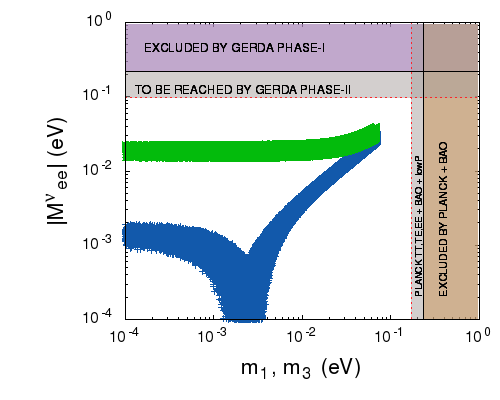}\\
\includegraphics[scale=.4]{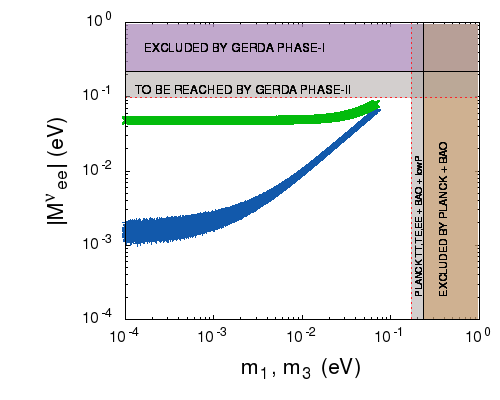} \includegraphics[scale=.4]{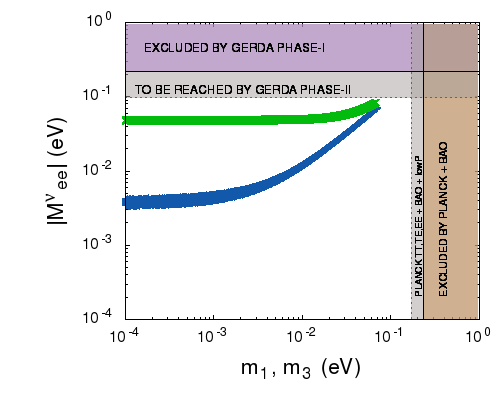}
\caption{Plot of  $|M^\nu_{ee}|$ vs. the lightest neutrino mass: the top two figures  represent Case A (left) and Case B (right) while the figures in the lower panel represent Case C (left) and Case D (right).}\label{f2}
\end{figure}

In our model there are four sets of values of the CP-violating phases $\alpha$ and $\beta$ for each neutrino mass ordering . Since $|M^\nu_{ee}|$ is sensitive to  the CP phases, we get four different plots for each mass ordering as shown in Fig. \ref{f2}. The same plots are valid for both types of mass ordering provided the horizontal axis is taken to represent the lightest neutrino mass $m_1$ or $m_3$ - depending on the ordering.
As mentioned earlier, we have used the upper  bound of 0.23 eV on $\Sigma_im_i$. These plots lead to upper bounds on the lightest neutrino mass for both cases of mass ordering. For hierarchical neutrinos,  $|M^\nu_{ee}|$ is found to lead to an upper limit which is below the  reach of the GERDA phase-II data. The latter appears close to being obtainable only for a quasidegenerate neutrino mass spectrum ($m_{lightest}> 0.07$ eV). However the value predicted in our model could  be probed by a combination of GERDA and MAJORANA experiments \cite{majo}. In order to explain the nature of the plots analytically, let us first consider the inverted mass ordering: In this case, with the approximations $m_3\simeq 0$ and $m_1 \simeq m_2$, the probed effective mass $|M^\nu_{ee}|$  simplifies to
\bea
|M^\nu_{ee}|=\sqrt{|\Delta m_{32}|^2}c_{13}^2 [\lbrace1-s_{12}^2(1-\cos \alpha)\rbrace^2+s_{12}^4\sin^2\alpha]^{1/2}.\label{ew}
\eea
 Clearly, $|M^\nu_{ee}|$ is insensitive to the  phases $\beta$ and $\delta$. On the other hand, for $\alpha=0$ and $\pi$ (\ref{ew}) simplifies to 
\bea
|M^\nu_{ee}| = \sqrt{|\Delta m_{32}|^2}c_{13}^2
\eea
and
\bea
|M^\nu_{ee}|=\sqrt{|\Delta m_{32}|^2}c_{13}^2 [\lbrace 1-2s_{12}^2\rbrace ^2 ]
\eea
respectively. Hence, for $\alpha=\pi$ (cases A, B ), $|M^\nu_{ee}|$ is suppressed as compared to the case $\alpha=0$ ( C, D). For a normal mass ordering, in addition to the $s_{13}$ suppression, there is a significant interference between the first two terms, thus lowering the value of $|M^\nu_{ee}|$. However, if $\alpha=0$, the first two terms interfere constructively  and then we obtain a lower bound ($\sim 10^{-3}$ eV for Case C and $\sim 5\times 10^{-3}$ eV for Case D) despite this being a case of  normal mass ordering. This is one of the remarkable results of the present analysis. On the other hand, for $\alpha=\pi$, the first two terms interfere destructively, for the case of a normal mass ordering; consequently, a sizable cancellation between them  brings down the value of $|M^\nu_{ee}|$ and results in the kinks shown by the lower curves in the top two figures.

\noindent
{\it CP asymmetry in neutrino oscillations }\\

Here we discuss the determination of our predicted maximal Dirac CP-violating phase $\delta$ by means of  neutrino oscillation studies. This $\delta$ will show up in  the asymmetry parameter $A_{\alpha\beta}$, defined as
\bea
A_{\alpha\beta}=P(\nu_{\alpha } \rightarrow \nu_{\beta})-P(\bar{\nu_{\alpha}}\rightarrow \bar{\nu_{\beta}}),
\eea 
where $\alpha,\beta =(e,\mu,\tau)$ are  flavor indices and the $P^,$s are transition probabilities. Let us consider first $\nu_{\mu}\rightarrow \nu_{e}$ oscillation in vacuum. The transition probability  can now be written (with the superscript zero indicating oscillations in vacuum) as
\bea
P_{\mu e}^0\equiv P^0(\nu_{\mu}\rightarrow \nu_{e})=P^0_{atm}+P^0_{sol}+2 \sqrt{ P^0_{atm}}\sqrt{ P^0_{sol}} \cos (\Delta_{32}+\delta),\label{nor}
\eea
where $\Delta_{ij}=\frac{\Delta m_{ij}^2L}{4E}$ is the kinematic phase factor ($L$ being the baseline length and $E$ being the beam energy) and $P^0_{atm},P^0_{sol}$ are respectively defined as 
\bea
  \sqrt{ P^0_{atm}}=\sin \theta_{23}\sin 2\theta_{13}\sin \Delta_{31},\\
  \sqrt{ P^0_{sol}}=\cos\theta_{23}\cos \theta_{13}\sin 2\theta_{12}\sin \Delta_{21}.
\eea
For an antineutrino beam, $\delta$ is replaced by $-\delta$ and thus we have
\bea
\bar{P}^0_{\mu e}\equiv P^0(\bar{\nu_{\mu}}\rightarrow \bar{\nu_{e}})=P^0_{atm}+P^0_{sol}+2 \sqrt{ P^0_{atm}}\sqrt{ P^0_{sol}} \cos (\Delta_{32}-\delta).\label{anti}
\eea
Now the CP asymmetry parameter $A^0_{\mu e}$  in vacuum \cite{cp} can be calculated as 
\bea
A^0_{\mu e}= \frac{P^0_{\mu e}-\bar{P}^0_{\mu e}}{P^0_{\mu e}+\bar{P}^0_{\mu e}}=\frac{2 \sqrt{ P^0_{atm}}\sqrt{ P^0_{sol}}\sin \Delta_{32}\sin\delta}{P^0_{atm}+P^0_{sol}+2 \sqrt{ P^0_{atm}}\sqrt{ P^0_{sol}}\cos\Delta_{32}\cos\delta}.\label{cpa}
\eea
With our prediction  $\cos\delta=0$, (\ref{cpa}) can be rewritten as
\bea
A^0_{\mu e}= \pm \frac{2 \sqrt{ P^0_{atm}}\sqrt{ P^0_{sol}}\sin \Delta_{32}}{P^0_{atm}+P^0_{sol}},
\eea
with a + ($-$) sign  for $\delta=\pi/2$ ($3\pi/2$).

 In order to realistically  describe neutrino oscillations in  long baseline experiments, matter effects in neutrino propagation through the earth need to be taken into account. In that case $P^0_{atm}$ and $P^0_{sol}$ will be modified to
\bea
\sqrt{P_{atm}}=\sin \theta_{23}\sin 2\theta_{13}\frac{\sin (\Delta_{31}-aL)}{\Delta_{31}-aL} \Delta_{31},\\
\sqrt{ P_{sol}}=\cos\theta_{23}\cos \theta_{13}\sin 2\theta_{12 } \frac{\sin aL}{aL}\sin \Delta_{21}
\eea
respectively. Here $a=G_F N_e/ \sqrt{2}$ with $G_F$ as the Fermi constant and $N_e$ as the number density of electrons in the medium of propagation. An approximate value of $a$ for the earth is 3500 $\rm km^{-1}$ \cite{cp}. Now the same formulae for $P_{\mu e}$, $\bar{P}_{\mu e}$ and $A_{\mu e}$ will hold as in (\ref{nor}), (\ref{anti}) and (\ref{cpa}) but with $P^0_{atm}$ and $P^0_{sol}$ replaced by $P_{atm}$ and $P_{sol}$ respectively.

In Fig.\ref{fig3} we plot $P_{\mu e}$ and $A_{\mu e}$  against the baseline length L in the two cases $\delta=\pi/2$ and $\delta=3\pi/2$ for both normal and inverted mass ordering. The lengths corresponding to T2K, $\rm NO\nu A$ and DUNE are  indicated in these figures. In Fig.\ref{fig4} the CP asymmetry $A_{\mu e}$ is plotted  against the beam energy E again for the cases $\delta=\pi/2$ and $\delta=3\pi/2$ separately for the three above cited experiments; both normal and inverted mass ordering cases are included. As expected, $A_{\mu e}$ has opposite signs for $\delta=\pi/2$ and $\delta=3\pi/2$. It is further interesting that the extrema of the CP-asymmetry parameter exhibit opposite behavior as a function of E for $\delta=\pi/2$ and $\delta=3\pi/2$.
\section{Summary}\label{s6}
In this paper we have proposed a complex extension of the scaling ansatz for the neutrino Majorana mass matrix $M^\nu$. To that end, we have made use of the residual $\mathbb{Z}_2\times \mathbb{Z}_2^{scaling}$ symmetry of $M^\nu$ by obtaining the representation $G_3^{scaling}$ from the original simple scaling ansatz on $M^\nu$. The resultant form of the neutrino Majorana matrix is given by $M_\nu^{CES}$ of (\ref{e3}). We have shown that it admits nonzero values of all the physical neutrino masses as well as both normal and inverted types of mass ordering. We have shown how a nonvanishing $\theta_{13}$ emerges from $M_\nu^{CES}$. The additional result $k^{-1}=\tan \theta_{23}$, $k$ being the real positive scaling factor, has also been derived. Dirac CP-violation has been shown to be maximal with $\cos\delta=0$ while Majorana CP-violation has been demonstrated to be absent with $\alpha,\beta=0$ or $\pi$. The type-I seesaw mechanism which yields nonzero neutrino masses within our scheme has also been constructed.  Phenomenological implications for both $0\nu\beta\beta$ decay and neutrino/antineutrino oscillation studies at long baselines have been worked out and projections made that will be testable in forthcoming experiments.

\appendix 
\section{Appendix: Derivation of the results on CP-violation even with the inclusion of the unphysical matrix}  \label{A1}
As  mentioned in Sec.\ref{s1},  our calculations have been done in a Weak Basis where the unphysical phases are absorbed in the neutrino fields. However, one can also reproduce our results including this unphysical phase matrix in the calculation. In that case the $U_{PMNS}$ of (\ref{eu}) writes as
\bea
U_{PMNS}=P_\phi U,\label{a1}
\eea
with $P_\phi=$ diag. ($e^{i\phi},1,1$). Note that there is only a single unphysical phase in the phase matrix $P_\phi$, since the symmetry under consideration dictates $M_\nu^{CES}$ in (\ref{e3}) to contain seven real parameters which correspond to three nonzero masses, three mixing angles and an unphysical phase. Now for $\tilde{d}_1=-1$, Eq. (\ref{e27}) and (1,1) element of the $U_{PMNS}$ in (\ref{a1}) gives
\bea 
e^{-2i\phi}=1,
\eea
therefore, $\phi=0$ or $\pi$. From (1,2) element we get 
\bea
e^{-i(\alpha+2\phi)}=-\eta.\label{a2}
\eea
Thus for both the values of $\phi$, (\ref{a2}) leads to (\ref{30k}); therefore, for each $\tilde{d}$ matrix with $\tilde{d}_1=-1$, the prediction for $\alpha$, i.e, $\alpha=0$ or $\pi$ remain the same. Now, following the same way as  in Sec.{\ref{s4}}, the results presented in Table \ref{t2} can be reproduced.\\

Unlike the previous case, now $\tilde{d}_1=1$ cannot be ruled out. In this case, from the (1,1) element of $U_{PMNS}$ in (\ref{a1}), we get $\phi=\pi/2$ or $3\pi/2$. Now, for both the values of $\phi$, Eq. (\ref{a2}) with $\eta=1$ leads  to $\alpha=0$, and with $\eta=-1$ leads to $\alpha=\pi$. Since the predictions for $\alpha$ remain the same, so do the other parameters which are solved exactly in the same way as in Sec.\ref{s4}, by use of {\it both} real and imaginary parts of  the relevant complex equations. We put the more general statements regarding the CP phases for each $\tilde{d}$ with $\tilde{d}_1=1$ in Table \ref{at}. In comparison with Table \ref{t2}, the values 

 \begin{figure}[H]
\hspace{3cm} N  \hspace{8.2cm} N \\
\includegraphics[scale=.45]{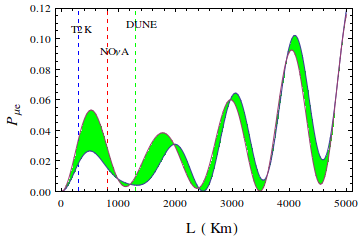} \hspace{3cm}\includegraphics[scale=.45]{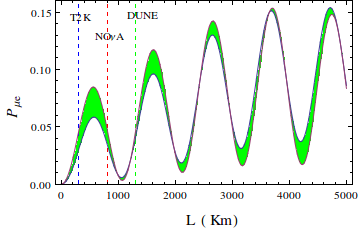}\\

\hspace{3cm} I  \hspace{8.2cm} I \\
\includegraphics[scale=.45]{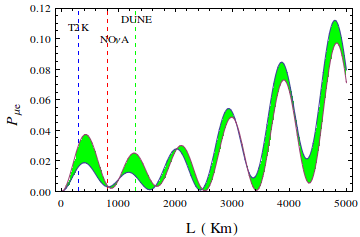} \hspace{3cm}\includegraphics[scale=.45]{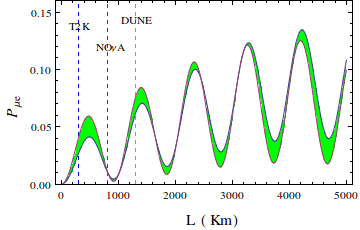}\\

\hspace{3cm} N  \hspace{8.2cm} N \\
\includegraphics[scale=.45]{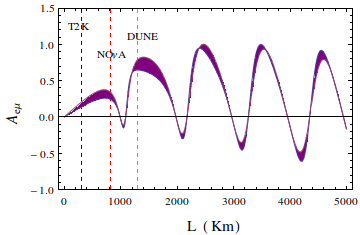}\hspace{3cm} \includegraphics[scale=.45]{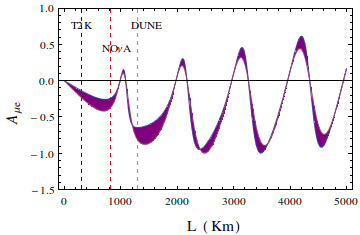}\\

\hspace{3cm} I  \hspace{8.2cm} I \\
\includegraphics[scale=.45]{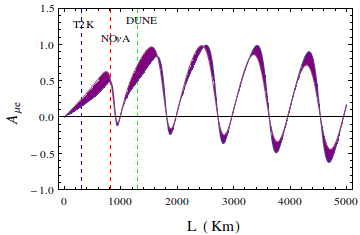} \hspace{3cm}\includegraphics[scale=.45]{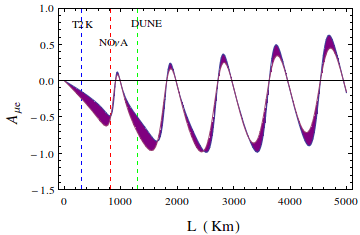}\\
\caption{Plots of the transition probability ($P_{\mu e}$) and CP asymmetry parameter ($A_{\mu e}$)  with baseline length L for $\delta=\pi/2$ (left panel) and $\delta=3\pi/2$ (right panel) with E = 1 GeV. Cases for normal (inverted) mass ordering have been labelled on top by N (I). The bands are caused by the atmospheric mixing angle $\theta_{23}$ being allowed to vary within the $3\sigma$ region while the other parameters are kept at their best fit values. }\label{fig3}
\end{figure}
\begin{figure}[H]
\hspace{3cm} N  \hspace{8.2cm} N \\
 \includegraphics[scale=.45]{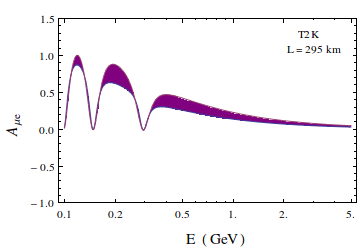} \hspace{3cm}\includegraphics[scale=.45]{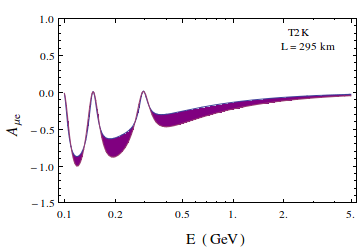}\\
 
 \hspace{3cm} I  \hspace{8.2cm} I \\
 \includegraphics[scale=.45]{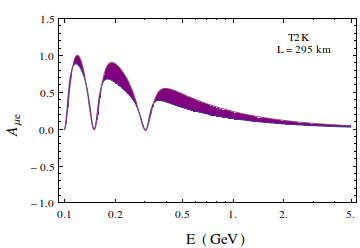} \hspace{3cm}\includegraphics[scale=.45]{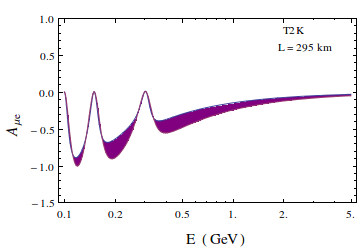}\\
 
 \hspace{3cm} N  \hspace{8.2cm} N \\
\includegraphics[scale=.45]{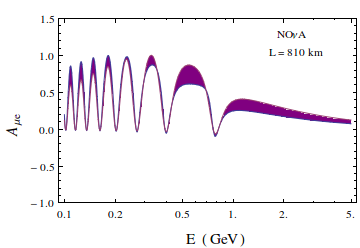} \hspace{3cm}\includegraphics[scale=.45]{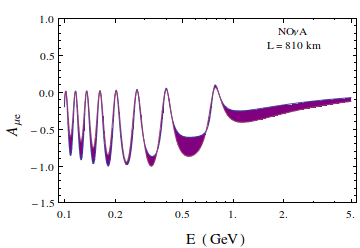}\\

\hspace{3cm} I  \hspace{8.2cm} I \\
\includegraphics[scale=.45]{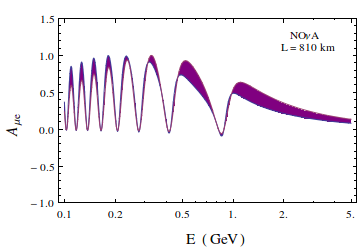} \hspace{3cm}\includegraphics[scale=.45]{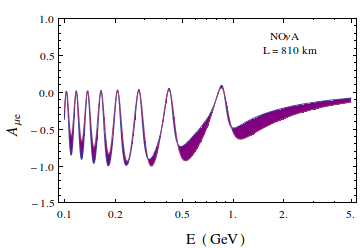}\\

 \end{figure}
 \hspace{12cm} (Fig. 4 contd.)
\begin{figure}[H]
 \hspace{3cm} N  \hspace{8.2cm} N \\
\includegraphics[scale=.45]{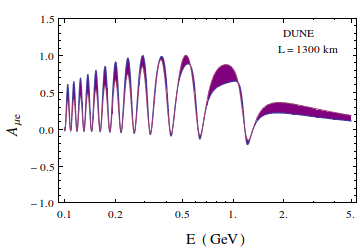} \hspace{3cm}\includegraphics[scale=.45]{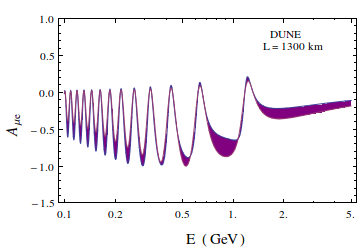}\\

\hspace{3cm} I  \hspace{8.2cm} I \\
\includegraphics[scale=.45]{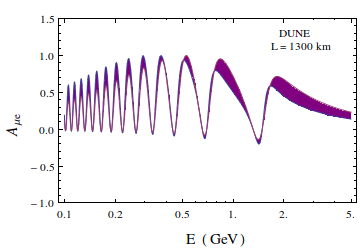} \hspace{3cm}\includegraphics[scale=.45]{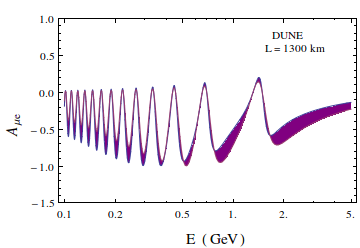}
 \caption{Plots of the CP asymmetry parameter $A_{\mu e}$  against beam energy E for $\delta=\pi/2$ (left panel) and $\delta=3\pi/2$ (right panel) for various experiments as shown. Cases for normal (inverted) mass ordering have been labelled on top by N (I). The atmospheric mixing angle $\theta_{23}$ is allowed to vary within the $3\sigma$ region, leading to the bands, while the other parameters are kept at their best fit values. }\label{fig4}
 \end{figure} 

\begin{table}[h]
\begin{center}
\caption{Predictions of the CP phases for $\tilde{d}_1=1$} \label{at}
 \begin{tabular}{|c|c|c|c|} 
\hline 
$\tilde{d}$&$\alpha$ &$\beta$ &$ \cos \delta$ \\
\hline
$\tilde{d}_e=\rm diag \hspace{1mm}(-1,+1,+1)$&$0$&$0$&$0$\\
$\tilde{d}_f=\rm diag \hspace{1mm}(-1,+1,-1)$&$0$&$\pi$&$0$\\
$\tilde{d}_g=\rm diag \hspace{1mm}(-1,-1,+1)$&$\pi$&$0$&$0$\\
$\tilde{d}_h=\rm diag \hspace{1mm}(-1,-1,-1)$&$\pi$&$\pi$&$0$\\
\hline 
\end{tabular} 
\end{center} 
\end{table} 
\noindent 
of $\alpha$ have changed relative to those of $\beta$, but the final result that both $\alpha$ and $\beta$ are either 0 or $\pi$ remain the same, though the value of $\tilde{d}_1$ has changed. 

\section*{Acknowledgements}
We thank P.S Bhupal Dev and Ketan M. Patel for their comments. RS and AG acknowledge the Department of Atomic Energy, Govt. of India, for financial support.  The work of PR has been supported by Indian National Science Academy.

{}

\begin{thebibliography}{}
\bibitem{King:2015aea} 
  S.~F.~King,
  J.\ Phys.\ G {\bf 42}, 123001 (2015),
  S.~Verma,
  Adv.\ High Energy Phys.\  {\bf 2015}, 385968 (2015).
\bibitem{Hagedorn:2016lva} 
  M. Fukugita and T. Yanagida, Phys. Lett. B174, 45 (1986). S. Davidson, E. Nardi and Y. Nir, Phys. Rept. 486, 105 (2008). C.~Hagedorn and E.~Molinaro,
  arXiv:1602.04206 [hep-ph].
\bibitem{Abe:2015awa} 
  K.~Abe {\it et al.} [T2K Collaboration],
  Phys.\ Rev.\ D {\bf 91}, no. 7, 072010 (2015).
\bibitem{Bian:2015opa} 
  J.~Bian [NOvA Collaboration],
  arXiv:1510.05708 [hep-ex].
  \bibitem{skk} 
 URL  \texttt{ https://indico.cern.ch/event/361123/session/2/contribution/348/attachments/1136004/}\\\texttt{ 1625868/SK atmospheric \_ kachulis\_ dpf2015.pdf}
\bibitem{Gonzalez-Garcia:2015qrr} 
  M.~C.~Gonzalez-Garcia, M.~Maltoni and T.~Schwetz,
  arXiv:1512.06856 [hep-ph].
\bibitem{Dell'Oro:2015oey} 
  S.~Dell'Oro, S.~Marcocci and F.~Vissani,
  Proceedings, 16th International Workshop on Neutrino Telescopes (Neutel 2015), PoS NEUTEL {\bf 2015}, 069 (2015).
  \bibitem{branco}
  G.~C.~Branco, L.~Lavoura and J.~P.~Silva, {\bf CP Violation},
(Clarendon Press,  Oxford, 1999).
\bibitem{PDG}
K.A Olive et al [Particle Data Group] Chinese Physics C38, 090001 (2014).
\bibitem{models}
L. Lavoura, Phys. Rev. D {\bf 62}, 093011 (2000). R.N Mohapatra and W. Rodejohann, Phys. Lett. B {\bf 644}, 59 (2007). 
\bibitem{scaling}
M.S. Berger and S. Santana, Phys. Rev. D {\bf 74}, 113007 (2006). A. Blum, R.N. Mohapatra and W. Rodejohann, Phys. Rev. D {\bf 76}, 053003 (2007). A.S Joshipura and W. Rodejohann, Phys. Lett. B {\bf 678}, 276(2009). B. Adhikary. M. Chakraborty and A. Ghosal, Phys. Rev. D {\bf 86}, 013015 (2012). M. Chakraborty, H.Z. Devi and A.Ghosal, Phys. Lett. B {\bf 741}, 210 (2015). A. Ghosal and R. Samanta, JHEP {\bf 1505}, 077 (2015).  R.~Samanta, M.~Chakraborty and A.~Ghosal, Nucl.\ Phys.\ B {\bf 904}, 86 (2016). 
\bibitem{th13} 
B.Z Hu et al. (Daya Bay Collaboration), Phys. Rev. Lett. {\bf 115}, 111802 (2015) and references therein. 
\bibitem{Lam}
C.S Lam, Phys. Lett B {\bf 656}, 193 (2007); Phys. Rev. Lett. {\bf 101}, 121602 (2008); Phys. Rev. D {\bf 78}, 073015 (2008).
\bibitem{ding}
 P.~Chen, G.~J.~Ding, F.~Gonzalez-Canales and J.~W.~F.~Valle,
  Phys.\ Lett.\ B {\bf 753}, 644 (2016).
\bibitem{mutau}
W. Grimus and L. Lavoura, Acta Phys. Polon. B {\bf 34}, 5393 (2003); Phys. Lett. B {579}, 113 (2004); Fortsch. der Phys. {\bf 61}, 535 (2013).
\bibitem{Grimus:2004cc} 
  W.~Grimus, A.~S.~Joshipura, S.~Kaneko, L.~Lavoura, H.~Sawanaka and M.~Tanimoto,
  Nucl.\ Phys.\ B {\bf 713}, 151 (2005).
\bibitem{dicus}
S.F Ge, D.A. Dicus and W. W Repko,  Phys.\ Rev.\ D {\bf 83}, 093007 (2011);  Phys. Lett. B {\bf 702}, 220 (2011); Phys. Rev Lett. {\bf 108}, 041801 (2012).
\bibitem{gencp}
G. Ecker, W. Grimus, H. Neufeld, J.Phys. A20, L807 (1987); Int.J.Mod.Phys. A3, 603 (1988).
  W.~Grimus and M.~N.~Rebelo,
  Phys.\ Rept.\  {\bf 281}, 239 (1997).
  S.~Gupta, A.~S.~Joshipura and K.~M.~Patel,
  Phys.\ Rev.\ D {\bf 85}, 031903 (2012).
G.~J.~Ding, S.~F.~King and A.~J.~Stuart,
  JHEP {\bf 1312}, 006 (2013). 
  C.~Hagedorn, A.~Meroni and E.~Molinaro,
  Nucl.\ Phys.\ B {\bf 891}, 499 (2015).
 P.~Chen, C.~Y.~Yao and G.~J.~Ding,
  Phys.\ Rev.\ D {\bf 92}, no. 7, 073002 (2015).
\bibitem{chen}
  P.~Chen, G.~J.~Ding and S.~F.~King,
  JHEP {\bf 1603}, 206 (2016).
\bibitem{Feruglio:2012cw} 
  F.~Feruglio, C.~Hagedorn and R.~Ziegler,
  JHEP {\bf 1307}, 027 (2013).

\bibitem{minakata}
 H. Minakata, H. Nunokawa and A. A. Quiroga, PTEP
2015
, 033B03 (2015).
\bibitem{planck} 
  P.~A.~R.~Ade {\it et al.} [Planck Collaboration],
  arXiv:1502.01589 [astro-ph.CO].
  
\bibitem{beta} 
  W.~Rodejohann,
  Int.\ J.\ Mod.\ Phys.\ E {\bf 20}, 1833 (2011),
  P.~S.~Bhupal Dev, S.~Goswami, M.~Mitra and W.~Rodejohann,
  Phys.\ Rev.\ D {\bf 88}, 091301 (2013).
\bibitem{kam} 
  K.~Asakura {\it et al.} [KamLAND-Zen Collaboration],
  Nucl.\ Phys.\ A {\bf 946}, 171 (2016).
\bibitem{exo} 
  M.~Auger {\it et al.} [EXO-200 Collaboration],
  Phys.\ Rev.\ Lett.\  {\bf 109}, 032505 (2012).

\bibitem{gerda1} 
  M.~Agostini {\it et al.} [GERDA Collaboration],
  Phys.\ Rev.\ Lett.\  {\bf 111}, no. 12, 122503 (2013)
  doi:10.1103/PhysRevLett.111.122503
  [arXiv:1307.4720 [nucl-ex]].
\bibitem{gerda2} 
  B.~Majorovits [GERDA Collaboration],
  AIP Conf.\ Proc.\  {\bf 1672}, 110003 (2015)
  [arXiv:1506.00415 [hep-ex]].
\bibitem{majo} 
  N.~Abgrall {\it et al.} [Majorana Collaboration],
  Adv.\ High Energy Phys.\  {\bf 2014}, 365432 (2014).
  \bibitem{cp}
 H. Nunokawa, S. Parke and J. W. Valle, CP violation and neutrino oscillations,
Progress in Particle and Nuclear Physics
60
(2008) 2 338-{ 402}.


\end{thebibliography}
\end{document}